\documentclass[a4paper]{jpconf}

\usepackage{cite}
\usepackage{subfigure}
\usepackage{amsmath,amssymb,amsfonts}
\usepackage{algorithmic}
\usepackage{graphicx}
\usepackage{textcomp}
\usepackage{xcolor}

\begin{document}

\title{Setup and Characteristics of a Timing Reference Signal with sub-ps Accuracy for AWAKE}

\author{Fabian Batsch for the AWAKE Collaboration}

\address{CERN, CH-1211 Geneva 23, Switzerland}

\ead{fabian.batsch@cern.ch}

\begin{abstract}
We describe a method to overcome the triggering jitter of a streak camera to obtain less noisy images of a self-modulated proton bunch over long time scales ($\sim$\,400\,ps to ns) with the time resolution ($\sim$\,1\,ps) of the short time scale images (73\,ps). We also determine that this method, using a reference laser pulse with a variable delay, leads to the determination of the time delay between the ionizing laser pulse and the reference pulse with an error of 0.6\,ps (rms).
\end{abstract}

\section{Introduction}
\label{intro}
\begin{figure}[h]
\centerline{\includegraphics[width=12cm]{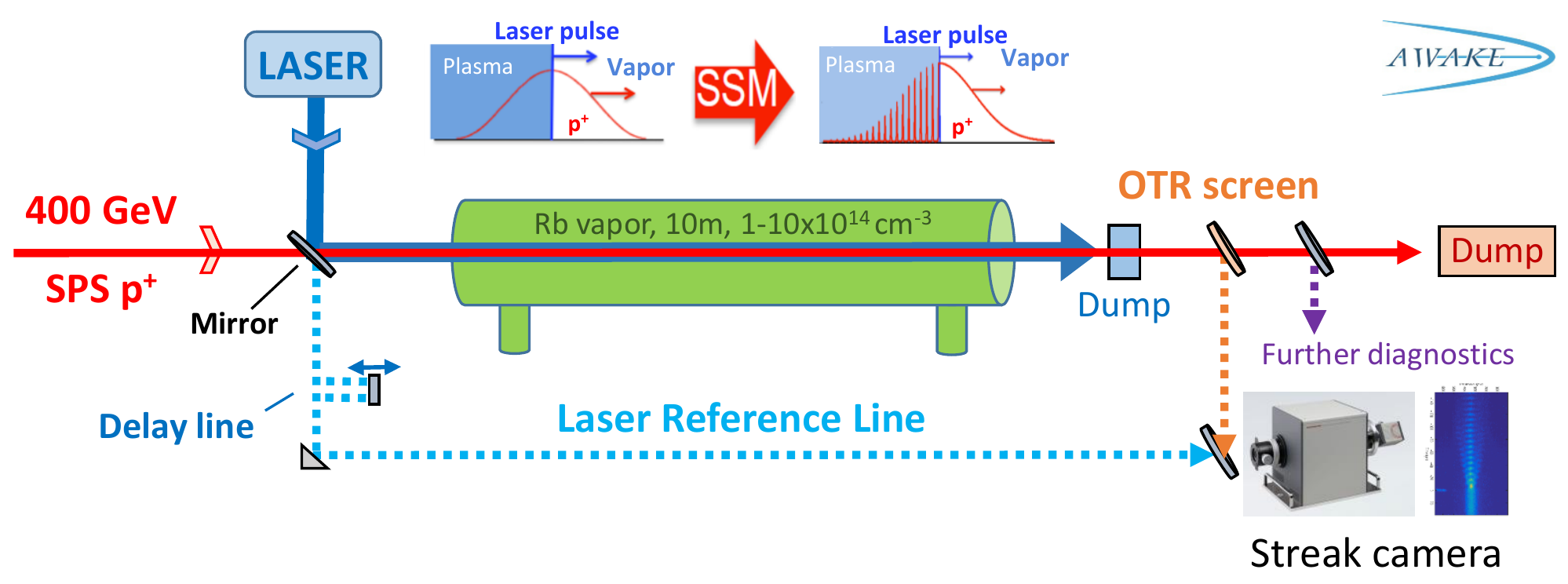}}
\caption{Schematic of the AWAKE experiment showing the different beam and diagnostic components, in particular the beam path of the laser timing reference signal.}
\label{fig:AWAKESetup}
\end{figure}

The AWAKE experiment is a proof-of-principle experiment at CERN demonstrating electron acceleration in a proton-driven plasma wakefield accelerator \cite{Nature, Statusreport2018}. A 10\,m-long vapor source provides rubidium (Rb) vapor with homogeneous temperature and density uniformity (better 0.2 $\%$ \cite{bib:oz}) (see Fig. \ref{fig:AWAKESetup}).
A Ti:Sapphire laser system ionizes the vapor (first e$^{-}$ of each Rb atom) which forms a plasma with $\sim$\,1 mm radius along the source. A co-propagating proton bunch from the CERN SPS (energy 400\,GeV, bunch length $\sigma_{z}=$\,6 -- 12\,cm) self-modulates into micro-bunches\cite{bib:karl} (see sketch in Fig. \ref{fig:AWAKESetup}). These micro-bunches resonantly drive wakefields, in which electrons from a 19\,MeV electron source are captured and accelerated. The self-modulation (SM) process can start from noise (so-called self-modulation instability, SMI) or can be seeded by placing the ionizing laser pulse within the proton bunch (seeded self-modulation, SSM) \cite{bib:karl, smi}. In AWAKE, the sharp ionization front serves as the seed.\\
Downstream of the vapor source, diverse beam diagnostics and the electron spectrometer are located. A thin metallic dump directly after the plasma blocks the ionizing laser pulse to protect these. The diagnostic to image the modulated proton bunch uses optical transition radiation (OTR) and a streak camera \cite{bib:karl}. The micro-bunches pass through a metallic foil, emitting OTR. This light propagates over a 15\,m-long transport line to a streak camera. \\
The streak camera triggering system exhibits a $\approx$ 5\,ps jitter (rms) with respect to the ionizing laser pulse (see chapter \ref{triggjitter}). For best time resolution ($\sim$ 1\,ps), the streak camera window is 73\,ps and the modulation period is in the 3 to 10\,ps range. Comparison of the relative modulation timing between events and over the long timescale of the proton bunch ($\sim$ 1\,ns) is therefore not possible. We thus developed and implemented a laser timing reference signal indicating the laser time-of-arrival on the streak camera images with sub-ps accuracy, as described in the following. 

\section{Experimental implementation}
\subsection{Setup and working principle}
\label{setup}
The chirp-pulse amplification Ti:Sapphire laser system produces pulses with an energy up to 450\,mJ and a duration of 120\,fs. For the timing reference signal, we pick up the bleed-through of the ionizing laser pulse from the first mirror after the laser system \cite{Statusreport2018}. This ensures that changes in the laser timing, originating from the amplifiers and the compressor in the laser chain, do not affect the synchronization between the reference signal and the ionizing laser pulse. An optical transfer line with in total 18 partially motorized, 2-inch-diameter mirrors and 1-inch periscopes guides the reference signal in free-space to the streak camera. It uses a separate, 60\,m-long path parallel to the proton beam line (partially using the vacuum transport pipe described in \cite{bib:florence}). Neutral density filters reduce the light intensity to protect the streak camera. The simplified path of the reference signal is sketched in Fig. \ref{fig:AWAKESetup} as the dashed blue line. It includes a delay line, consisting of a  motorized translation stage with 15\,cm travel range and a retro reflector. It allows for remotely adjusting the time-of-arrival of the reference signal on the streak camera over the range of 1\,ns. The free-space propagation of the reference signal provides a dispersion-minimized transport and a pulse length comparable to thus of the ionizing laser pulse.\\ 
\begin{figure}[b]
\centerline{\includegraphics[width=16cm]{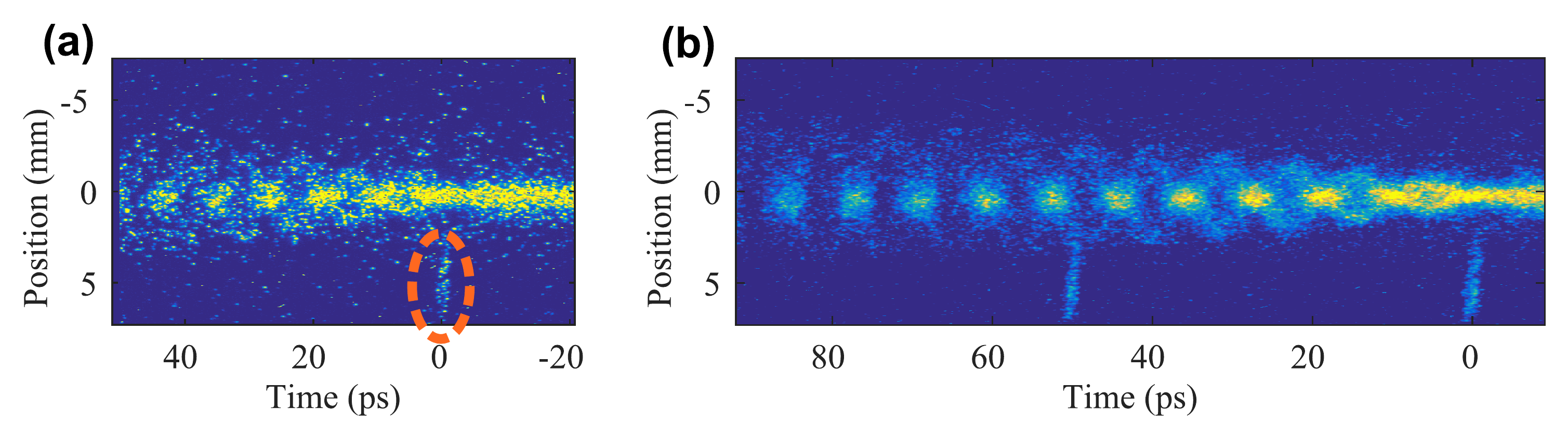}}
\caption{(a) Streak camera image showing the self-modulation behind the ionization laser pulse and the laser reference signal, marked by the orange circle. The head of the bunch is at negative time values. The event was recorded using the 73\,ps streak camera window and with a Rb vapor density of 1.8$~\times$~10$^{14}$\,cm$^{-3}$. (b) Reconstructed image of the self-modulated proton bunch. We delayed the reference signal by 50\,ps with respect to the zero timing and averaged over nine events for each delay position of the reference signal. }
\label{fig:streakimage}
\end{figure}
 The streak camera ($512\times672$\,pixel) time-resolves the spatial structure of the incoming OTR light with time windows from 50\,ns to 73\,ps. Figure \ref{fig:streakimage}(a) shows a single streak camera image of a self-modulated proton bunch behind the ionization point, as recorded on the 73\,ps window. The reference signal, visible at the bottom of the image at $t=0$\,ps (marked by a circle), shows the time-of-arrival of the ionization front. Its timing is used as the zero point for most SM analysis, including the studies on seeding processes, the phase reproducibility of the SM and on the influence of a plasma gradient on the SM, all described in separate publications and to be published soon. For image processing, series of images with similar experimental parameters can be synchronized in time with respect to the reference signal timing and numerically added together or averaged, despite the triggering system jitter. The resulting image of the proton bunch is an average of the events with improved signal-to-noise ratio and reveals more details of the SM process.\\
 To obtain an averaged image of the entire proton bunch using the highest time resolution window (73\,ps) and thus analyse the micro-bunches 10s to 100s of ps behind the seed point, we introduce a delay between the reference signal and the ionization front using the translation stage (see Fig.\ref{fig:AWAKESetup}). We delay the trigger of the streak camera by the same amount. 
 Then, we stitch together these superimposed events according to the delay of the reference signal. Figure \ref{fig:streakimage}(b) shows an example for the resulting procedure of superimposing and stitching events. Nine consecutive events are superimposed for each of the two delay steps. The reference signal was delayed once by 50\,ps. Repeating this procedure several times allows for imaging the entire self-modulated proton bunch with the highest resolution of 1\,ps and the improved signal-to-noise ratio.
 
\subsection{Time matching}
\label{timematching}
To indicate the time-of-arrival of the ionizing laser with sub-ps accuracy, the path length of the reference signal line and of the ionizing laser beam line ($\approx\,60\,$m) must be matched with an accuracy better $300\,\mu$m (or 1\,ps). To verify this, we send the ionizing laser pulse in low energy mode along the beam path onto the OTR foil and observe both pulses simultaneously on the streak camera. In high laser energy mode, the ionizing laser pulse is blocked by an aluminium foil placed between the plasma and the OTR foil and is not visible on streak camera images (e.g. Fig.\ref{fig:streakimage}(a)). We use the delay line to match the reference signal and ionizing laser time-of-arrival and thus their path lengths. For the matching, one has to take into account the different propagation times through optics material for the low-energy laser beam at 780\,nm and the broadband OTR light (filtered to ($450\pm 10)$\,nm to maximize the time resolution of the streak camera \cite{bib:karl}). The optical dispersion induced by optical elements in the beam path between the OTR light and the laser light was measured to be $9.9\,$ps \cite{bib:martyprivate}. It can be compensated for with the delay line to obtain a reference signal effectively synchronized with OTR light and thus with the SM. Any error on this delay would correspond to a systematic error in timing and would not influence measurements between events. \\

\section{Characteristics}
\begin{figure}[ht]
\centerline{\includegraphics[height=5.6cm]{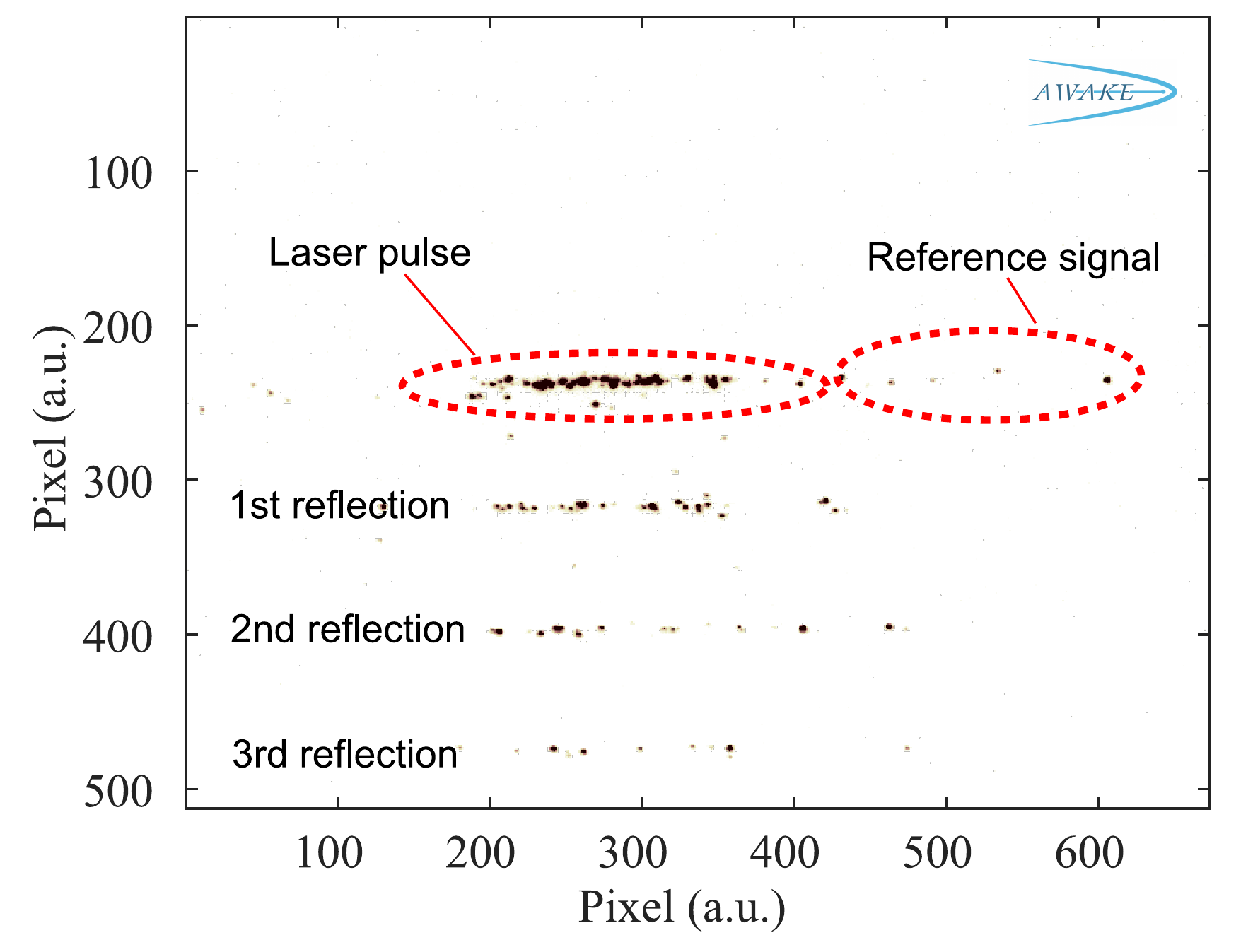}}
\caption{Streak camera image showing the laser pulse and its three reflections. }
\label{fig:markerref}
\end{figure}

The purpose of the laser timing reference signal is to be used as a timing reference with respect to the ionizing laser pulse, with a variable delay. 
During experiments, the measured variation in relative time is a convolution of the real time variation and the measurement uncertainties from the timing reference signal. These main uncertainties originate from a timing jitter between the ionizing laser pulse and the reference signal (caused by pointing jitter, mechanical vibrations and air draft), from the uncertainty in determining the  time-of-arrival of the reference signal and from the error on the translation stage moving distance. 
To characterize these uncertainties, we set the reference line delay so that it is visible together with the laser in low energy mode on the same streak camera image with a 73\,ps window. For this test measurement, we also introduce a 5\,mm-thick piece of glass in the beam path to obtain multiple surface reflections. Figure \ref{fig:markerref} shows a streak camera image with the reference signal as well as the low power ionizing laser pulse including its reflections.

\subsection{Trigger system jitter between streak camera and laser}
\label{triggjitter}
\begin{figure}[ht]
\centerline{\includegraphics[height=5.5cm]{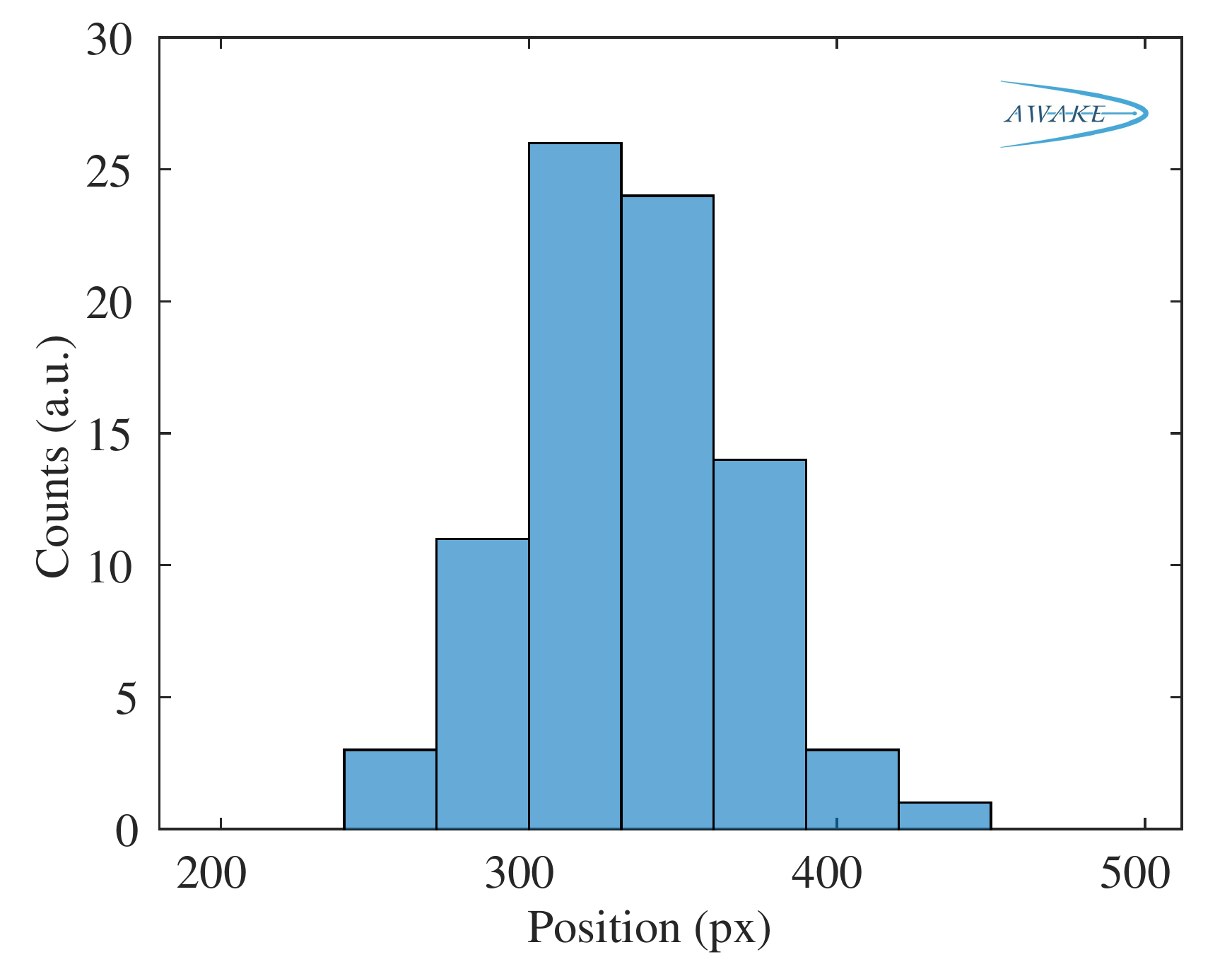}}
\caption{Histogram of the detected reference signal position in pixel on the 73\,ps window. The variations are due to the triggering system jitter.}
\label{fig:laswindojitter}
\end{figure}

Before characterizing the measurement errors, we quantity the trigger system jitter of the streak camera time window relative to the laser reference signal by measuring the pixel (time) at which the reference signal appears on the streak camera image for fixed settings. Figure \ref{fig:laswindojitter} shows the distribution of the signal. The laser reference signal jitter covers a total range of 160\,pixel or 22.8\,ps, with a standard deviation of 34\,pixel = 4.8\,ps. This jitter does not impact directly the data analysis thanks to the use of the reference signal, but might cause the signal and region of interest on the streak camera image to jitter off the screen.

\subsection{Uncertainty on the temporal position of the reference signal}
\label{uncertTempPos}
\begin{figure}[t]
\centerline{\includegraphics[height=5.6cm]{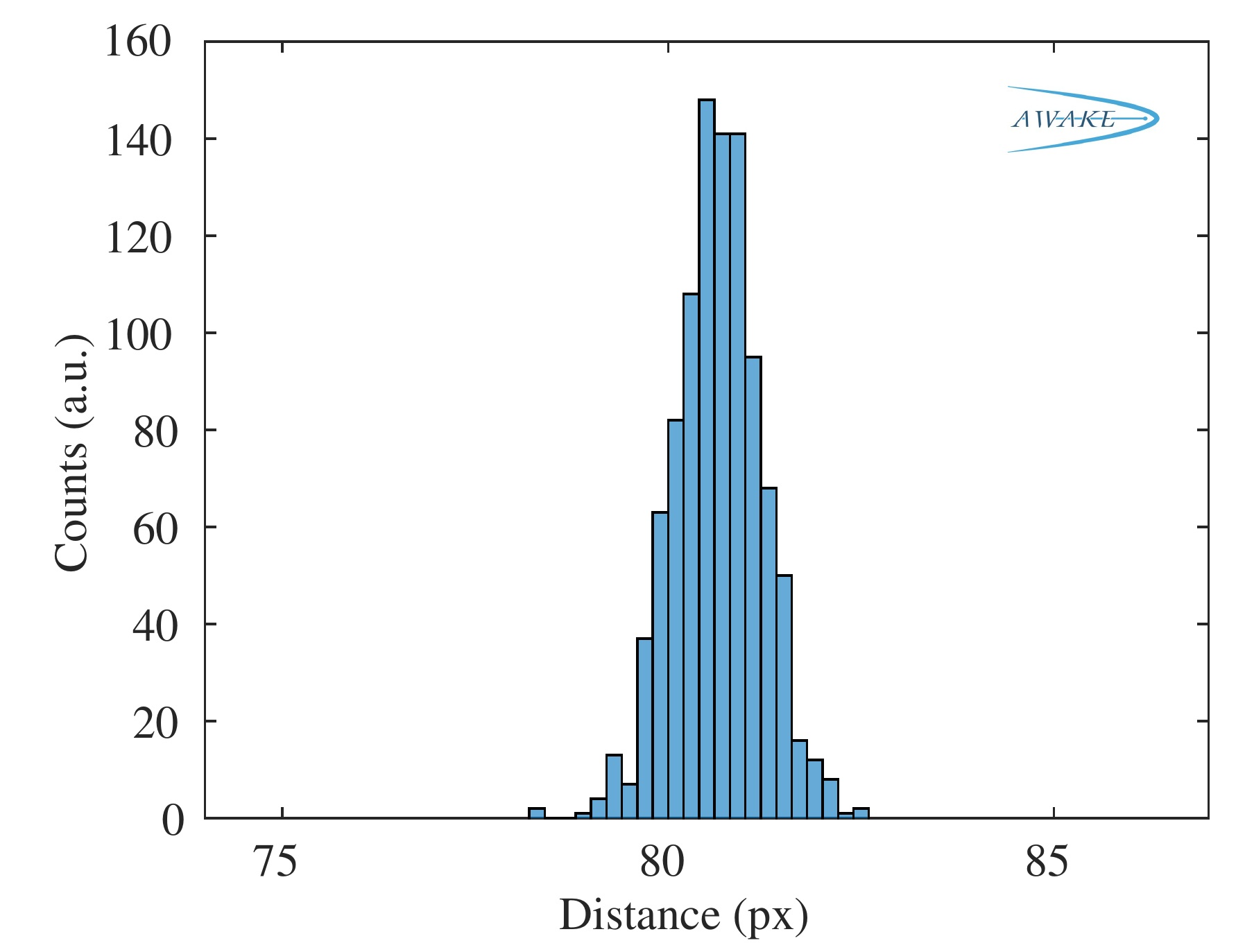}}
\caption{Detected temporal spacing between the laser light and its reflection, observed with the streak camera. The measured mean value is $(80.6 \pm0.6)\,$pixel.}
\label{fig:markeraccubbb}
\end{figure}

The most likely contribution to the measurement error is the determination of the temporal position of the reference signal on the streak camera images. From these images, we determine the timing by fitting a Gaussian profile to a line-out obtained by summing the signal over 50\,pixel (vertical direction in Fig.\ref{fig:streakimage}(a)) that only contains the reference signal. 
For the test here (Fig.\ref{fig:markerref}), having the multiple laser pulse reflections is similar to having several reference signals. It allows for determining the uncertainty of the algorithm by measuring the spacing between two reflections. Hereby we assume that for different events, the temporal spacing between two reflections does not change since the propagation distances of the reflected light is unaltered. From the uncertainty of this difference, we can then estimate the uncertainty in determining the position in time of the reference signal. Figure \ref{fig:markeraccubbb} shows the distribution of the detected distance between the main signal and the first reflection, measured in pixels. 
The histogram has a rms width (equal to two standard deviations) of $1.1\,$pixel. On the 73\,ps streak window, this corresponds to an uncertainty in measuring their spacing of 157\,fs. The uncertainty for the determination of a single reference pulse is accordingly $157/\sqrt{2}=111$\,fs. However, this value is only relevant for events with well adjusted intensity of the reference signal as in this test measurement. When a stronger signal is recorded, the signal broadens in time inside the streak camera due to space charge effects and thus increases the uncertainty. We avoid this by increasing the strength of the filters that attenuate the intensity of the reference signal (described in chapter \ref{setup}).

\subsection{Timing jitter between laser pulse and reference signal}
\label{lasermarkerjitter}
The pointing jitter of the ionizing laser pulse, mechanical vibrations or air drafts inside the transport line of the reference signal might cause a relative timing jitter between the ionizing laser pulse and the reference signal. To quantify this timing jitter, we determine and compare their relative positions from 200 streak camera images as shown in Fig.\ref{fig:markerref}. The difference in their positions shows a standard deviation of 1\,pixel, equal to 143\,fs on the streak camera window. Since this value is 
convoluted with and smaller than the uncertainty in determining the timing of both pulses of 157\,fs (see chapter \ref{uncertTempPos}), we conclude that no measurable jitter is observed. 

\subsection{Precision of the reference signal delay setting}
For the analysis of the SM in the back of the proton bunch (i.e.\ with delays with respect to the ionizing laser pulse larger than the streak camera window), the precision and reproducibility with which the timing reference signal delay is set gains importance. The displacement of the translation stage in the delay line is set in units of mm. The corresponding temporal offset is twice the travel distance divided by the speed of light. 
To test the reliability of the translation stage settings and thus of the reference signal delay, we move the stage step-wise over a total travel range of 60.0\,mm (the range we use in the measurements) and measure repeatably the actual displacement $x_m$ versus the set distance $x_{set}$ using a caliper with an accuracy of 0.05\,mm (corresponds to 0.08\,\% over 60\,mm). 
\begin{figure}[ht]
\centerline{\includegraphics[height=5.6cm]{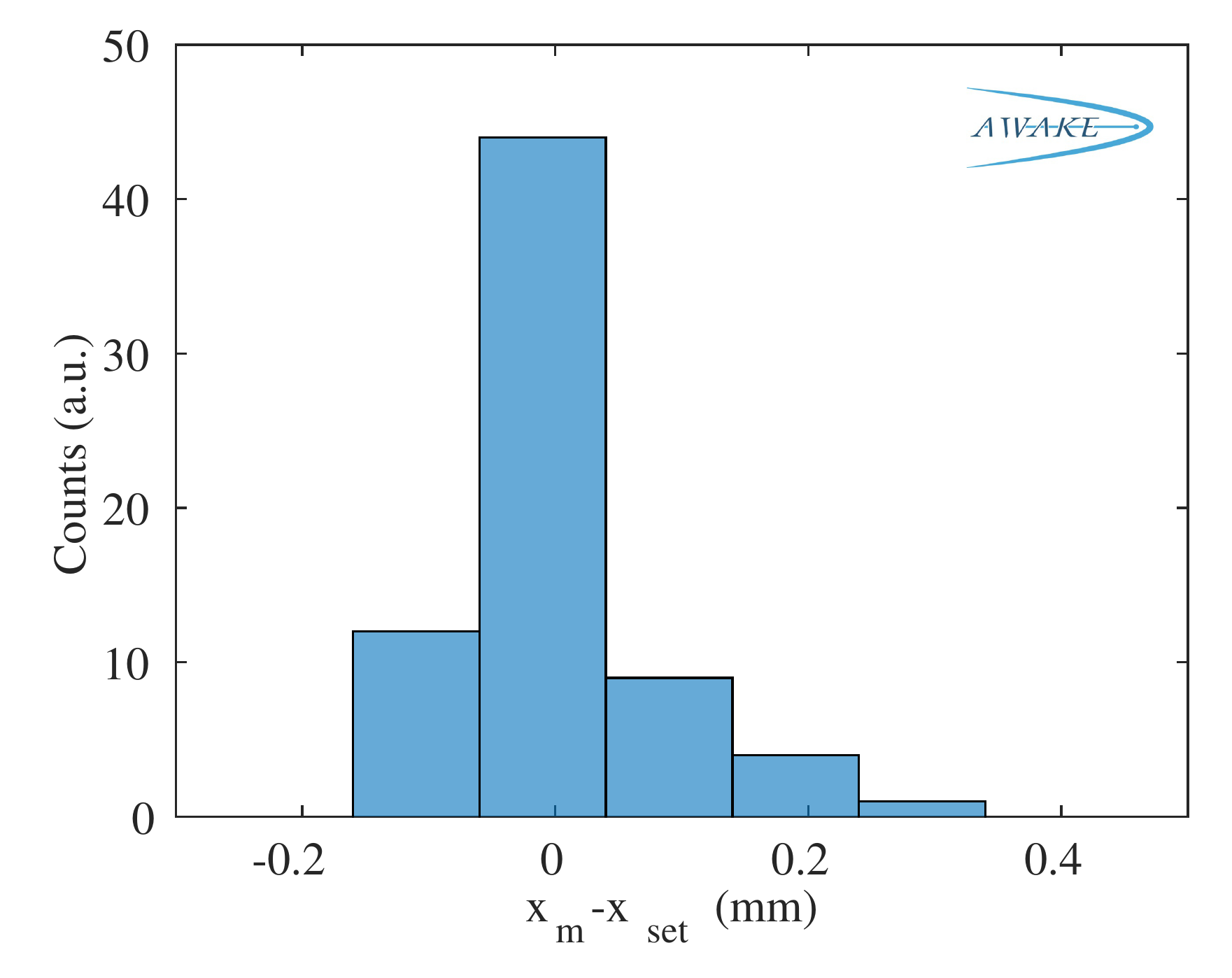}}
\caption{Histogram of the deviation between measured and set displacement of the translation stage.}
\label{fig:stagehisto}
\end{figure}
 The histogram in Fig.\ref{fig:stagehisto} shows the deviation of the measured from the set distance. The distribution has a rms width of 0.16\,mm. Thus, the corresponding error in setting the delay of the reference signal is 529\,fs (rms).

\subsection{Total uncertainty}
We measured the three main uncertainties that the laser timing reference signal impose on the analysis of the self-modulation of a proton bunch: the uncertainty in determining the temporal position of the reference signal on streak camera images (111\,fs rms), the jitter of the relative delay between both beams (below measurement limit) and an error introduced by the uncertainty in the translation stage position when changing the delay of the reference signal with respect to the laser pulse (529\,fs). To combine these uncertainties, one has to take into account that the uncertainty on the temporal position on the streak camera must be considered twice when determining the delay between the reference signal and laser pulse with the same procedure. 
The resulting total error (added in quadrature) is 552\,fs (rms), with the largest contribution from setting the delay with the translation stage.

\section{Summary}
We describe the implementation of a laser timing reference signal into the AWAKE experiment by propagating a replica of the ionizing laser pulse in a free-space transfer line to the streak camera diagnostic. We show the procedure to determine and adjust the delay between this reference signal and the ionizing laser pulse. This allows for denoised, high-resolution ($\sim$\,1\,ps) images of self-modulated proton bunches over time scales much longer (100s of ps) than the streak camera window (73\,ps width) despite a triggering system jitter between the streak camera and the ionizing laser pulse. 
We measure this jitter using the reference signal to be 4.8\,ps (rms). We characterize the error this reference signal implies on data analysis. The uncertainties that originate from determining the timing of the reference signal on the streak camera images, from the jitter between main and reference signal and from setting the delay with respect to the ionizing laser pulse add up to a total uncertainty of 0.6\,ps (rms).

\section*{Acknowledgment}
This work is sponsored by the Wolfgang Gentner Program of the German Federal Ministry of Education and Research (05E15CHA). 

\nocite{bib:florence}

\end{document}